\documentclass[epj,final]{svjour}
\usepackage[numbers,sort&compress]{natbib}
\usepackage{amssymb,amsmath,array}
\usepackage{graphicx}
\usepackage{subfigure}

\makeatletter
 
\newcommand{\bs}[1]{\boldsymbol{#1}}
\newcommand{\vc}[1]{\mathbf{#1}}
\newcommand{\uvc}[1]{\mathbf{\hat #1}}

\newcommand{\dd}{\mathrm{d}}
\newcommand{\pdr}[1]{\frac{\partial #1}{\partial t}}
\newcommand{\drf}[1]{\frac{\dd #1}{\dd t}}
\newcommand{\trans}{\emph{trans} }
\newcommand{\cis}{\emph{cis} }

\renewcommand{\Re}{\mathop{\rm Re}\nolimits}

\makeatother

\begin{document}
\title{Spatial reorientation of azobenzene side groups of a liquid
  crystalline polymer induced with linearly polarized light}

\author{%
Yu.~Zakrevskyy\inst{1} \and
O.~Yaroshchuk\inst{1} \and 
A.D.~Kiselev\inst{2} \and  
J.~Stumpe\inst{3} \and J. Lindau\inst{4}
}

\institute{%
Institute of Physics of NASU, pr. Nauki 46,
03028 Ky\"{\i}v, Ukraine \and  
Chernigov State Technological University, Shevchenko
Street 95, 14027 Chernigov, Ukraine\\
\email{kisel@elit.chernigov.ua}
 \and   
Fraunhofer Institute of Applied Polymer Research,
Erieseering 42, 10319 Berlin, Germany \and 
Institute of Phys. Chemistry, Martin-Luther University, M\"{u}hlphorte 1, 
06106 Halle, Germany
}

\titlerunning{Spatial reorientation in LC azopolymer}
\authorrunning{Yu.~Zakrevskyy, O.~Yaroshchuk,
A.D.~Kiselev \textit{et al.}} 

\mail{A.D.~Kiselev}

\abstract{%
  The photoinduced 3D orientational structures in films of a liquid
  crystalline polyester, containing azobenzene side groups, are
  studied both experimentally and theoretically.  By using the null
  ellipsometry and the UV/Vis absorption spectroscopy, the
  preferential in-plane alignment of the azobenzene fragments and
  in-plane reorientation under irradiation with linearly polarized UV
  light are established.  The uniaxial and biaxial orientational order
  of the azobenzene chromophores are detected.  The biaxiality is
  observed in the intermediate stages of irradiation, whereas the
  uniaxial structure is maintained in the photosaturated state of the
  photoorientation process.  The components of the order parameter
  tensor of the azobenzene fragments are estimated for the initial
  state and after different doses of irradiation.  The proposed theory
  takes into account biaxiality of the induced structures.  Numerical
  analysis of the master equations for the order parameter tensor is
  found to yield the results that are in good agreement with the
  experimental dependencies of the order parameter components on the
  illumination time. 
}

\date{} 
\maketitle

\section{Introduction}
\label{sec:intro}

Effect of photoinduced anisotropy (POA) implies that the optical
anisotropy revealed itself as  dichroism of absorption or
birefringence is brought about in medium under the action of light.
The capability of having the light-controlled anisotropy makes the
materials that exhibit POA very promising and highly perspective for
use in many photonic applications such as optical data storage and
processing, telecommunication and reversible
holography~\cite{Pras:1995,Eich:1987}.  In addition, it was found that
substances with POA effect serve as excellent aligning substrates for
liquid crystals~\cite{Gib:1991,Ichim:1988}.

Polymers containing covalently linked photo\-chro\-mic moieties such
as azobenzene derivatives are known as azo\-po\-lymers.  These
materials exhibit POA of extremely high efficiency: the value of
photoinduced birefringence in azo\-po\-ly\-mers can be as high as 0.3
and the dichroic ratio of the absorption is over 10.  It makes
azopolymers particularly suitable for the investigation of light
induced ordering processes.  This is why in the last decade these
polymers have been the subject of intense experimental and theoretical
studies~\cite{Eich:1987,Hvil:1992,Wies:1992,Dum2:1993,Stu:1994,Ped:1997,
  Nat:1998,Puch:1998,Puch:1999}.
     
The accepted mechanism of POA induced by the linearly polarized UV
light involves induced \textit{trans--cis}-photo\-iso\-meri\-zation
and subsequent thermal and/or photochemical
\textit{cis--trans}-back-iso\-meri\-zation of the azobenzene moieties.
Since the optical dipole of $\pi\pi^*$ and of the n$\pi^*$ transition
of the azobenzene moiety is directed along its long molecular axis,
the fragments oriented perpendicular to the actinic light polarization
vector, $\vc{E}$, then become almost inactive, whereas the other with
suitable orientation are active undergoing photoisomerization. These
\textit{trans--cis--trans} photoisomerization cycles are accompanied
by rotations of the azobenzene chromophores resulting eventually in an
orientation of the long axes of the azobenzene fragments along all
directions normal to the polarization vector of the incident actinic
light. Non-photoactive groups then undergo reorientation due
to cooperative motion or dipole
interaction~\cite{Nat:1998,Puch:1998,Puch:1999,Stu1:1994,Stu2:1994}.
 
The above scenario, known as photoorientation mechanism, assumes
angular redistribution of the long axes of the \trans molecules during
the \textit{trans--cis--trans} isomerization cycles.  It was initially
suggested in~\cite{Nep:1963} for the case, when the \cis state has a
short lifetime, that it becomes temporary populated during this
process but reacts immediately back to thermodynamically stable \trans
isomeric form.

Another limiting case is known as an angular selective hole burning
mechanism (photoselection) and occurs when the \cis states are long
living.  In this case POA is caused by the selective depletion of the
\trans isomeric form during the establishment of the steady
state~\cite{Dum2:1993}.  The anisotropy induced in this way is not
long term stable and disappears as a result of the thermal back
reaction. The photoorientation process in the steady state of the
photoisomerization takes place simultaneously, but it needs longer
time to saturate.  Generally, both mechanisms contribute to POA.
 
It is evident from the foregoing that the actinic light results in the
orientation of azobenzene chromophores perpendicular to the
polarization vector $\vc{E}$.  These directions can be thought as
equivalent provided that the symmetry group of the system includes
rotations.  From the experimental results, however, the latter is not
the case.  In particular, it was found that the photoinduced
orientational structures can show
biaxiality~\cite{Wies:1992,Yar:abs:1999,Kis:cond,Yar:2000}.  The
variety of orientational configurations (uniaxial, biaxial, splayed)
with different spatial orientations of the principle axes can be
expected depending on many factors such as chemical structure of
polymer, method of film preparation, irradiation conditions and so on.

In the past years this spatial character of the photoorientation has not
received much attention.  It was neglected in the bulk of experimental
and theoretical studies of POA in azobenzene containing
polymers~\cite{Eich:1987,Hvil:1992,Nat:1998,Dum2:1993,Stu:1994,
Ped:1997,Puch:1998}.  One conceivable reason for this can be the lack
of appropriate experimental methods.  On the other hand, until
recently, the problems related to the 3D orientational structures in
polymeric films has not been of major interest for applications.
But such kind of studies are currently of considerable importance in
the development of new compensation films for liquid crystal (LC)
displays~\cite{Witt:1997} and the pretilt angle generation by the use
of photoalignment method of LC orientation~\cite{Dyad:1995}.
 
The known methods suitable for the experimental study of the 3D
orientational distributions in polymer films can be divided into two
groups. 

The methods of the first group are based on absorption
measurements. These methods have the indisputable advantage that the
order parameters of various molecular groups can be estimated from the
results of these measurements.  Shortcomings of the known absorption
methods~\cite{Wies:1992,Blin:1983} are the limited field of
applications and the strong approximations.
 
The second group includes the methods 
dealing with principle refractive indices. 
Recently several variations of the prism 
coupling methods have been applied to measure the principle refractive 
indices in azopolymer films~\cite{Osm:1999,Feng:1995,Cim:1999}. 
These results, however,  were 
not used for in-depth analysis of such 
features of the spatial ordering as  
biaxiality  and spatial orientation of the optical axes depending 
on polymer chemical structure, irradiation conditions etc.
   
Our goal is a comprehensive investigation of the peculiarities of 3D 
orientational ordering in azopolymers. The present work is a part of 
the study focused on the orientational biaxiality and the 
transition from biaxial to uniaxial structures caused by the polarized 
actinic light.

The paper is organized as follows. 

In Sect.~\ref{sec:experiment} we describe our combined approach based
on using the methods that deal with both absorption and birefringence
measurements. The modified null ellipsometry method is employed to
study the general structure of the anisotropic polymer films.  The
components of order parameter tensor of the azobenzene chromophores
are estimated from the results of the UV absorption measurements.

Material of Sect.~\ref{sec:theory} comprises   
the theoretical part of the paper.
We begin with the analysis of general kinetic rate equations
and show how the known results~\cite{Ped:1997,Puch:1998}
can be recovered by using our theoretical approach.  
Then we formulate the
phenomenological model of the photoinduced ordering in azopolymers 
that accounts for  
biaxiality of the induced structures and  long term 
stability of POA. After computing the order parameter 
components of azobenzene units for different irradiation doses we 
find that the predictions of the theory are in good agreement with the 
data obtained experimentally.

Finally in Sect.~\ref{sec:concl} we draw together the results and make
some concluding remarks.

\section{Experimental}
\label{sec:experiment}

\subsection{Samples preparation and irradiation procedure}
\label{subsec:samples}

We investigated POA using
poly[octyl{(4-hexyloxy-4'-nitro)\-azo\-ben\-ze\-ne}malonate] 
as model polymer which
synthesis is described in~\cite{Bohm:1993}. 
The thermal properties of the polymer are characterized by
the transition temperatures
C$_1$-32$^o$-C$_2$-44$^o$-S 52$^o$-N-55$^o$-I 
detected by DSC and polarizing microscopy
of the polymer in the bulk. (Two crystalline states are
labelled C$_1$ and C$_2$; the symbols S and N stand for nematic and
smectic mesophases, respectively; the symbol I corresponds 
to the isotropic melt.) 
The polymer was solved in 
dichloroethane and spincoated on the quartz slabs. The prepared films 
were kept at the room temperature for 24~h for the evaporation of 
solvent. The thickness of the films of about 200--600~nm 
were measured with a profilometer of Tencor Instruments 
 
In order
to induce anisotropy in the films, we used the irradiation of a Hg 
lamp in combination with an interference filter (365~nm). 
The intensity of the actinic light was about 1.0~mW/cm$^{\,2}$. 
A Glan-Thompson polarizer was 
applied for the polarization of the UV light. A normal incidence of the 
actinic light was used in our studies. 
   
The irradiation was provided in several 
steps followed by both birefringence end absorption measurements.
In order to have ordering processes in the films completed after
switching off the irradiation, 
the waiting time interval before the measurements was longer than 15~min. 
According to~\cite{Puch:1999}, 
it corresponds to a time period 
which guarantee that all \cis isomers react back to the \trans form.

\begin{figure*}[!tbh]
\vskip10mm
\centering
\resizebox{180mm}{!}{\includegraphics{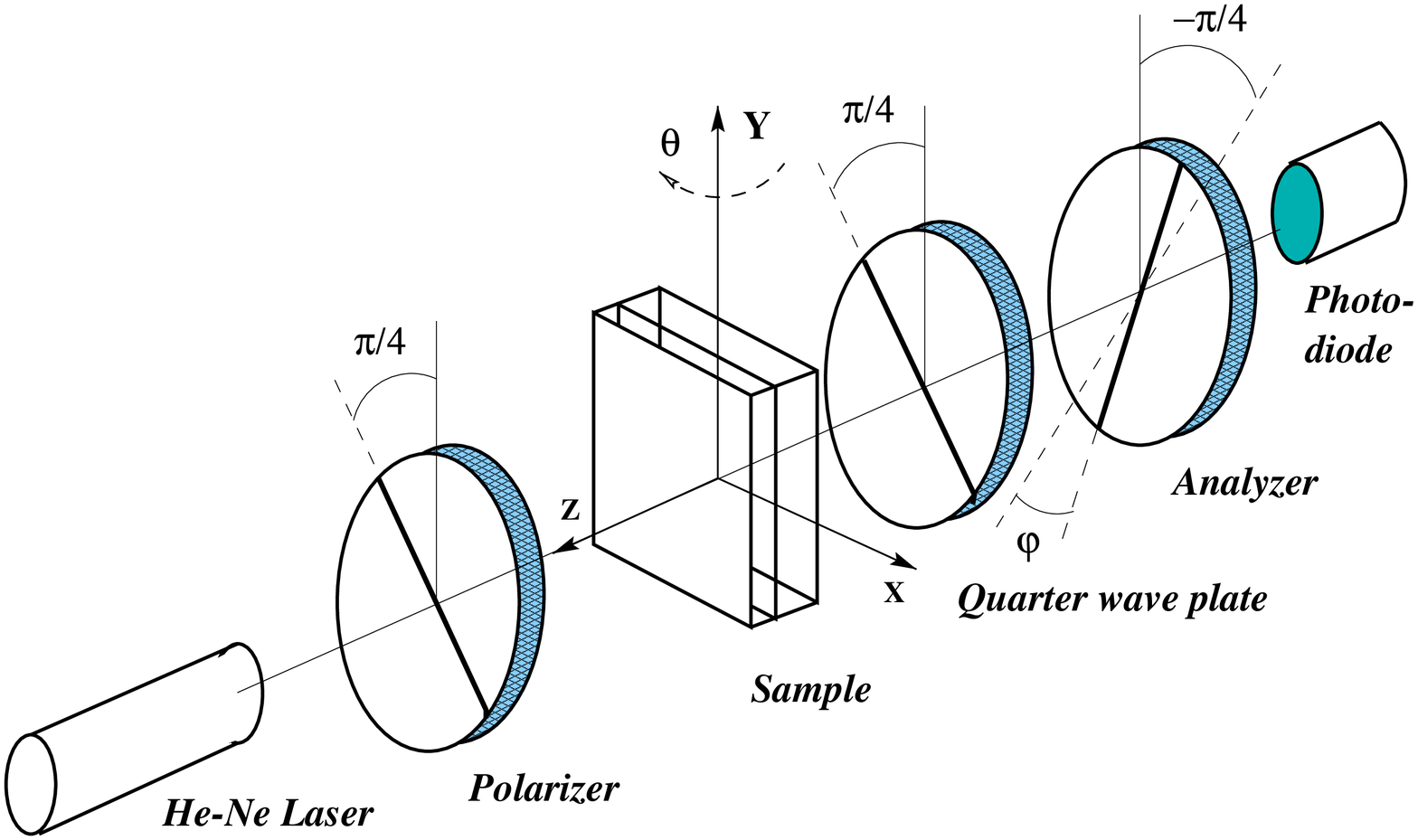}}
\caption{%
Experimental setup for the null ellipsometry measurements. 
}
\label{fig:1:exp}
\end{figure*}

\subsection{Null ellipsometry method}
\label{subsec:null-ell}

Instead of the prism coupling methods commonly used for the estimation 
of principle refractive indices we applied null ellipsometry 
technique~\cite{Azz:1977} dealing with birefringence components. 
By this means we have avoided 
some disadvantages of prism coupling method such as 
the problem of 
making optical contact between the prism and the polymer layer. 

The 
optical scheme of our method is presented in Fig.~\ref{fig:1:exp}. 
The polymer film 
is placed between crossed polarizer and analyzer and  a quarter wave 
plate with the optic axes oriented parallel to the polarization direction 
of the polarizer.  The elliptically 
polarized beam passed through 
the sample is transformed into the linearly polarized light
by means of the quarter wave plate. 
The polarization plane of this light is turned with respect to the 
polarization direction of the polarizer. 
This rotation is related to the 
phase retardation acquired by the light beam after passing through 
the film under investigation. 
It can be compensated by  
rotating the analyzer to the angle $\phi$ that encodes
information on the phase retardation.
 
This method used 
for the normal incidence of the testing light is known as the 
Senarmont technique. 
It is suitable for the in-plane birefringence 
measurements. 

Using oblique incidence of the testing beam
we have extended this method for estimation of both in-plane, 
$n_y-n_x$, 
and out-of-plane $n_z-n_x$ 
birefringence ($n_x$, $n_y$ and $n_z$ are 
the principle refractive indices of the film shown in 
Fig.~\ref{fig:1:exp}). 
In this case, the angle $\phi$  depends on the 
in-plane retardation $(n_y-n_x)d$, the out-of-plane retardation 
$(n_z-n_x)d$ 
and the absolute value of a refractive index of the biaxial 
film, say, $n_x$.

We need to have
the light coming out of the quarter wave plate  almost linear 
polarized when the system analyzes the phase shift between two 
orthogonal eigenmodes of the sample. 
In our experimental setup this requirement 
can be met, when the $x$ axis, directed along the 
polarization vector of the actinic light, is oriented horizontally or 
vertically. 
Dependencies of the analyzer rotation angle $\phi$ 
on the incidence angle of the testing beam $\theta$   
were measured for both vertical and 
horizontal orientation of the $x$ axis.
The value of $n_x$ was 
measured  with the Abbe refractometer independently. 
 
By using Berreman's 
$4\times 4$ 
matrix method~\cite{Berr:1972},
the $\theta$-dependencies of $\phi$ were calculated. 
Maxwell's equations for the light propagation 
through the system of polarizer, sample and quarter wave plate were 
solved numerically for the different configurations of optical axes in 
the samples.
The measured  and computed $\phi$ versus $\theta$   
curves were fitted in the most probable configuration model 
using the measured value of $n_x$. 

We conclude on alignment of the azobenzene fragments 
from the obtained values 
of $(n_y-n_x)d$  and $(n_z-n_x)d$ assuming that the preferred direction of 
these fragments coincides with the direction of the largest refractive 
index. More details on the method can be found in our 
previous publication~\cite{Yar:2000}. 

In our setup designed for the
null ellipsometry measurements we used a low power He-Ne laser 
($\lambda=632.8$~nm), 
two Glan-Thompson polarizers mounted on rotational sta\-ges 
from Oriel Corp., a quarter wave plate from Edmund Scientific and a 
sample holder mounted on the rotational stage. The light intensity was 
measured with a photodiode. The setup was automatically controlled by 
a personal computer. 
The rotation accuracy of the analyzer was better than 
0.2~degree.

\begin{figure*}[!tbh]
\vskip15mm
\centering
\resizebox{140mm}{!}{\includegraphics{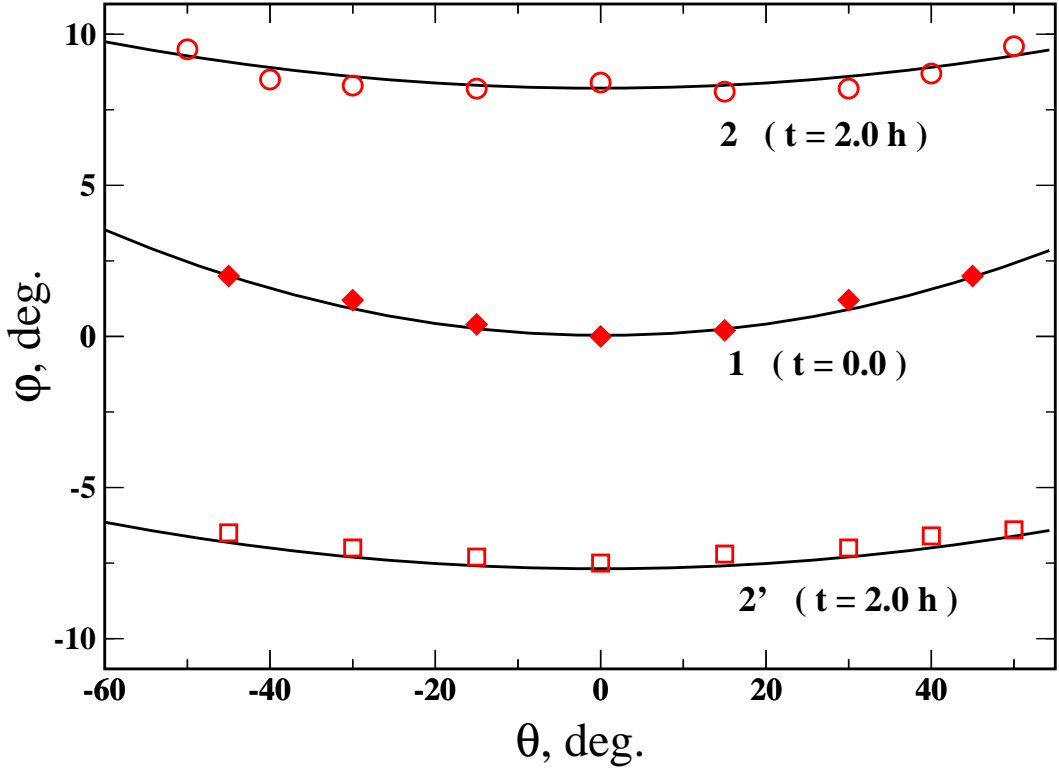}}
\caption{%
Dependencies of the phase shift $\phi$ on the incidence angle $\theta$ 
measured before (curve 1) and
after irradiation (curves 2 and 2'). 
}
\label{fig:2:exp}
\end{figure*}

\subsection{Absorption measurements}
\label{subsec:absorpt}

The UV/Vis absorption measurements were carried out using a diode 
array spectrometer (Polytec XDAP V2.3). The samples were set normally 
to the testing beam of a deuterium lamp. A Glan-Thompson prism with a 
computer-driven stepper was used for polarization of the testing beam. 
The UV spectra of the original as well as irradiated films were 
measured in the spectral range of 220--400~nm with the rotation step of 
polarizer of 5 degree.
 
From these measurements the optical density components 
corresponding to the absorption maximum of azobenzene chromophores were 
estimated for the polarization direction of the testing light parallel 
to $x$ and $y$ axes, respectively. We denote them as $D_x$  and $D_y$, 
respectively. The out-of-plane component, $D_z$, was estimated by making 
use the method proposed in~\cite{Wies:1992,Phaa:1996}.  
The latter assumes that the sample 
has uniaxial structure with in-plane position of the axis of 
anisotropy at the instant of time $t_0$.
It implies that $D_z(t_0)=D_x(t_0)$ and the 
total absorption can be estimated as follows
\begin{align}
  \label{eq:1}
  D_{\mathrm total}=D_x(t_0)+D_y(t_0)&+D_z(t_0)=\notag\\
&=2D_x(t_0)+D_y(t_0)\, .
\end{align}
When the number of \trans azobenzene units does not change
considerably, the total absorption is constant and the value 
of $D_z$ at instant of time $t$ can be determined from the 
following equation:
\begin{equation}
  \label{eq:2}
  D_z(t)=D_{\mathrm total}-D_x(t)-D_y(t)\, ,
\end{equation}
where $D_x(t)$ and $D_y(t)$ are the experimentally measured
parameters.

\begin{figure*}[!tbh]
\vskip15mm
\centering
\resizebox{140mm}{!}{\includegraphics{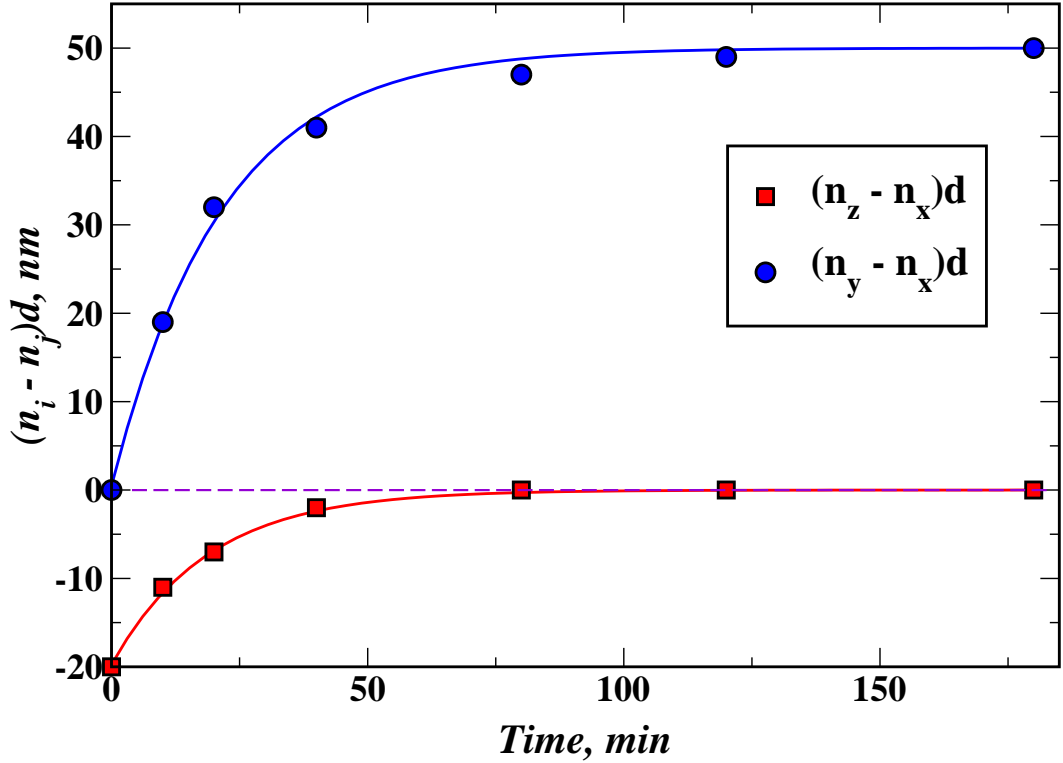}}
\caption{%
In-plane and out-of-plane birefringence as a function of the
irradiation time. 
}
\label{fig:3:exp}
\end{figure*}

\subsection{Experimental results}
\label{subsec:exp-res}

Figure~\ref{fig:2:exp} shows the phase shift $\phi$  
versus the incidence angle $\theta$  curve 
measured with null ellipsometry method for the 
non-irradiated polymer film. 
The curves $\phi(\theta)$ measured for vertical 
and horizontal position of $x$ axis overlaps. 
In addition, it is seen that 
there is no phase shift for normal light incidence ($\theta=0$). 
So we arrive at the conclusion
that the in-plane indices are matched: $n_y=n_x$.  
The film, however, possesses out of plane birefringence 
$(n_z-n_x)d=-20$~nm that results in 
a phase shift at oblique light incidence.

\begin{figure*}[!tbh]
\vskip15mm
\centering
\resizebox{140mm}{!}{\includegraphics{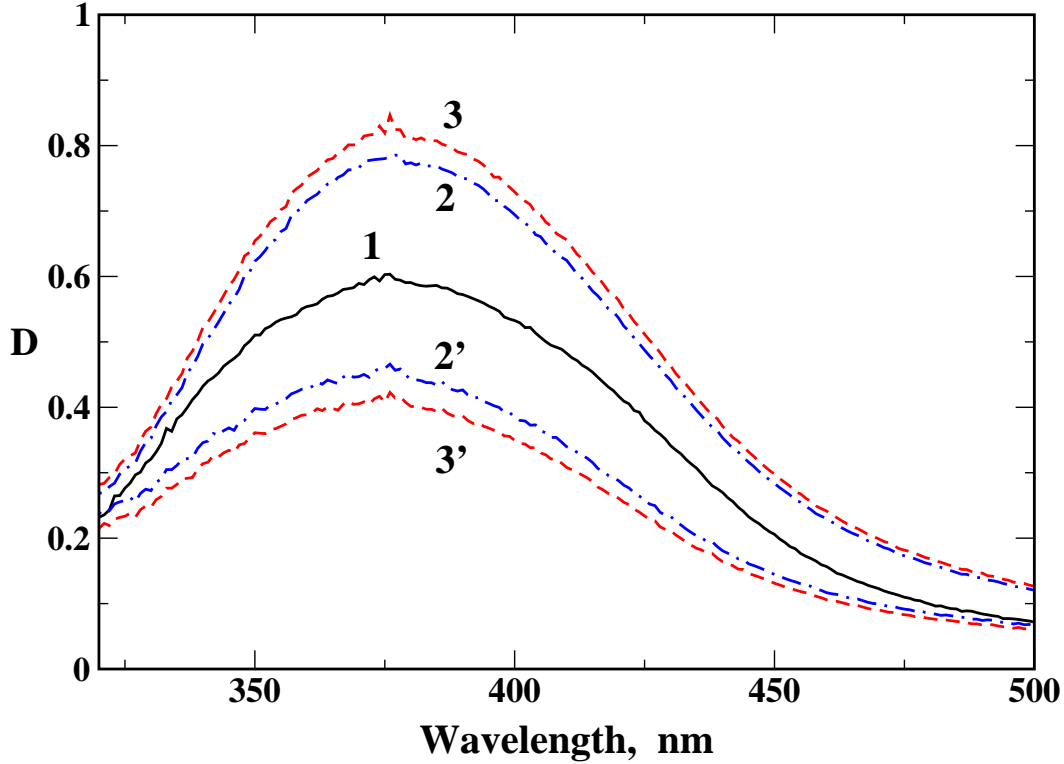}}
\caption{%
UV absorption spectra measured before (curve 1) and immediately
after the irradiation over 1~h (curves 2 and 2').
The stationary spectra $D_x$  and $D_y$ (the waiting time is 15 min) 
are shown as 
the curves 3 and 3', respectively. 
}
\label{fig:4:exp}
\end{figure*}

\begin{figure*}[!tbh]
\vskip5mm
\centering
\resizebox{130mm}{!}{\includegraphics{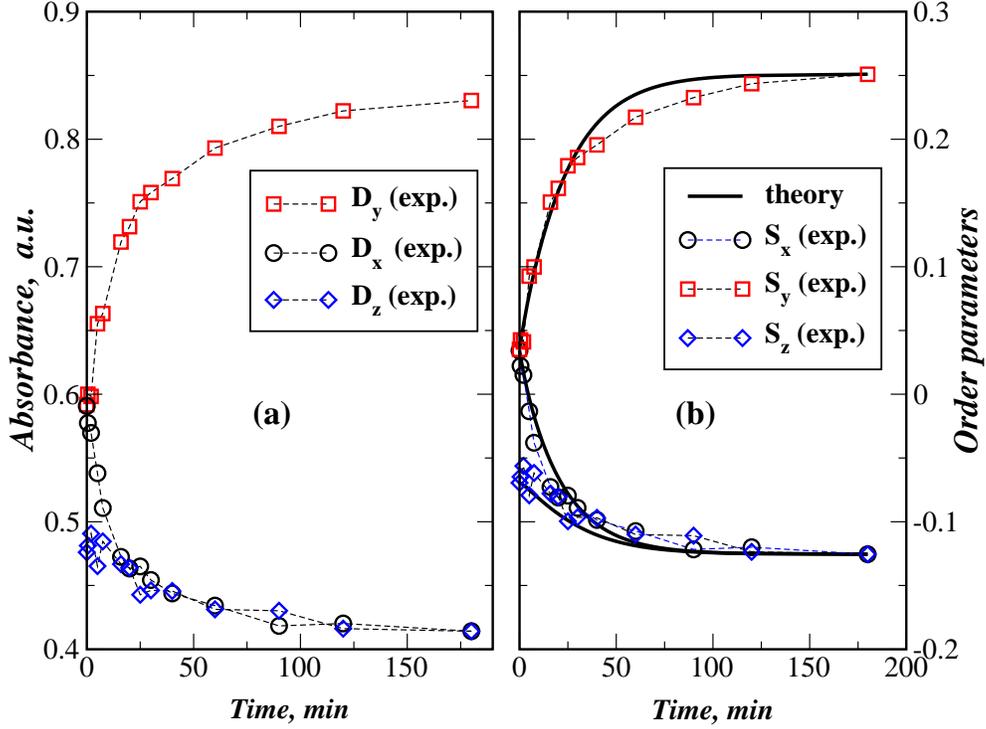}}
\caption{%
Dependencies of (a) the principle absorption coefficients and 
(b) the order parameter components on the irradiation time.
Theoretical curves for the diagonal components
of the order parameter tensor are shown as solid lines. 
}
\label{fig:1}
\end{figure*}

The fitting gives the 
following relations for the principle refractive indices: 
$n_z<n_x=n_y$ and $n_z-n_x\approx -0.1$. 
The film, therefore, demonstrates negative birefringence with the 
optical axis normal to the film surface. The relationship between 
the three indices suggests that the azobenzene fragments are randomly 
distributed in the plane of the film with no preferred direction for 
their orientation (a degenerate in-plane distribution).

Figure~\ref{fig:2:exp} shows the measured $\phi$ 
versus $\theta$  curves for the same polymer film 
after 2~h of UV light irradiation. Curves 1 and 2 correspond to 
vertical and horizontal position of the film $x$ axis, respectively.
 
According to the modeling, positive phase shift corresponds to the 
axis in the horizontal direction having the higher in-plane refractive 
index $n_y$ perpendicular to UV light polarization and the lower in-plane 
index $n_x$. From the curve fitting we have: $n_y-n_x=0.25$ 
($(n_y-n_x)d\approx 50 nm$), $(n_z-n_x)d=0$~nm, 
$n_y>n_x=n_z$. 
The light induced structure is positive uniaxial with the 
optical axis perpendicular to the UV light polarization. In this case, 
the azobenzene fragments show planar alignment perpendicular to the UV 
light polarization.

The fitted values of the in-plane, $n_y-n_x$, and out-of-plane
birefringence, $n_z-n_x$, for various irradiation times are presented
in Fig.~\ref{fig:3:exp}.  For small irradiation doses the principle
refractive indices are different: $n_z<n_x<n_y$.  The in-plane
birefringence monotonously increases up to the saturated value with
the increase of irradiation time.  On the other hand, the difference
between $n_x$ and $n_z$ decreases and becomes negligible in the
photosaturated state.  So the film is biaxial at the intermediate
stages of irradiation, whereas the photosaturated state can be
characterized as an uniaxial structure.  The optical axis of this
structure lies in the plane of the film and is directed along the $y$
axis.
 
The null ellipsometry identifies general
orientational structure, but it  does not provide the means to estimate 
the order parameters of various molecular groups. 
Different absorption methods are common for this purpose. 
In this case the 
wavelength of testing light is tuned to the absorption maximum of 
the selected molecular fragments. 

In order to estimate the order 
parameter of azobenzene units we carried out UV absorption 
measurements in the absorption maximum of azobenzene chromophores.
The UV/Vis spectrum of the studied azopolymer is presented in 
Fig.~\ref{fig:4:exp} (curve 1). 
It contains the intensive absorption band with the maximum at 
$\lambda=377$~nm corresponding to the $\pi\pi^{*}$ 
transition of \trans azobenzene fragments. 
 
The spectrum reveals polarization splitting during irradiation with
polarized light. The polarization components $D_x$ and $D_y$, measured
just after switching off the actinic light, are depicted in
Fig.~\ref{fig:4:exp} as 2 and 2', respectively. These spectra show
changes that become stationary for approximately 10~min.

The stationary spectra $D_x$  and $D_y$ are 
shown in Fig.~\ref{fig:4:exp}
as curves 3 and 3', respectively. In order to have the 
azobenzene units relaxed to the stationary state, the components $D_x$  
and $D_y$ were measured in 15~min after each irradiation period.
 
The experimentally measured absorption components $D_x$ and $D_y$ before 
irradiation and for different irradiation doses are presented in 
Fig.~\ref{fig:1}a. 
Kinetics of $D_x$  and $D_y$ is typical for reorientation mechanism of 
azobenzene units~\cite{Dum2:1993}.
Both curves reveal saturation. As it 
was shown by the null ellipsometry method, 
the saturated state of the polymer film under consideration 
is uniaxial with the in-plane 
orientation of the anisotropy axis. 

In order to prove that the method 
described in Sect.~\ref{subsec:absorpt}  
can be applied to estimate $D_z$, we need to 
show that, 
as compared to the non-irradiated film,
the number of \trans isomers does not change 
in the state relaxed after the irradiation. 
This is the case under 
the lifetime of \cis isomers is shorter than the 
time of spectral relaxation after switching off the actinic light.
 
In order to estimate the lifetime of \cis isomers we have measured
relaxation of the spectral changes at $\lambda_t=377$~nm after
irradiation with non-polarized light. Incidence directions of both
actinic and testing light were approximately normal to the film. It
was found that the relaxation curve $D(t)$ contains two components
with characteristic times of 1.0~sec and 4~min, respectively. The
first value can be attributed to \textit{cis-trans} transition of
azobenzene chromophores. 
The other time can be related to either the small fraction of
long living \cis isomers or more likely to the
orientational relaxation of the azobenzene units.
In any case, the waiting time of 15~min ensures that
we definitely have all the \cis
isomers transformed into the \trans form.  
 
So in our measurements the concentration of \trans isomers is preserved 
and the method described in Sect.~\ref{subsec:absorpt} 
can be applied. 
The values of $D_z$ 
calculated for various irradiation doses using 
Eqs.~\eqref{eq:1}--~\eqref{eq:2} are 
presented in Fig.~\ref{fig:1}a. 
Dependencies $D_x(t)$, $D_y(t)$  and $D_z(t)$ show that 
the photoinduced ordering is mainly due to the in-plane reorientation 
of azobenzene fragments in the $y$ direction. In addition, slight 
reorientation in the $z$ direction is observed.

The orientational structure is generally described 
by the tensor $S_{ij}$, 
which is diagonal when the coordinate axes 
directed along the principle axes of the film. 
The diagonal elements $S_{xx}\equiv S_x$, 
$S_{yy}\equiv S_y$ and $S_{zz}\equiv S_x$ are 
related to the absorption components $D_x$, $D_y$ and 
$D_z$~\cite{Wies:1992}. For example,
\begin{equation}
  \label{eq:3}
  S_{x}=\frac{2D_x-(D_y+D_z)}{2(D_x+D_y+D_z)}\, .
\end{equation}
The components $S_{y}$ and $S_{z}$ 
can be obtained by the cyclic 
permutation in the expression~\eqref{eq:3}.

The values of $S_{x}$,  $S_{y}$ and $S_{z}$ 
calculated using equation~\eqref{eq:3} are 
presented in Fig.~\ref{fig:1}b. 
As is seen, during the initial stage of 
irradiation orientational configuration is biaxial, 
whereas the initial and the photosaturated 
states are uniaxial. 
These structures are characterized by the 
following order parameters: 
$S_x=S_y\approx 0.035$ and $S_z\approx -0.07$
for the non-irradiated film;
$S_x=S_z\approx -0.1255$ and $S_y\approx 0.251$ 
for the film in the photosaturated state.

The transition between biaxial and uniaxial photoinduced orientations
will be the subject of subsequent studies. We believe that a tendency
to form an uniaxial structure is related to an intrinsic property of
self-organization of mesogenic groups.  Theoretically, 
similar tendency of nematic liquid crystals can be related to 
some specific features of the phase transition
reflected in the form of the mean field free energy~\cite{Patash}.
In this connection we assume
that the irradiation with polarized light causes both the
photoorientation of azochromophores and the self-aggre\-ga\-tion as it was
found for other liquid crystalline
azopolymers~\cite{Stu:1997a,Stu:1997b}.   

\section{Theory}
\label{sec:theory}

Before we proceed with theoretical considerations let us
emphasize the following distinguishing features of
POA in liquid crystalline azopolymers:
\renewcommand{\theenumi}{\alph{enumi}}
\renewcommand{\labelenumi}{(\theenumi)}
\begin{enumerate}
\item 
in contrast with the reversible POA, considered
in~\cite{Tod:1983,Dum:1992,Dum:1993,Dum:1996,Nep:1963}, where POA
disappears 
after switching off the irradiation, POA in liquid crystalline
azopolymers can only be thermally erased by heating above clearing
temperature.
The long term stability of POA is caused by the photoorientation
of the azobenzene; 

\item 
the biaxiality effects discussed in Sect.~\ref{sec:intro}.   
\end{enumerate}
Clearly, we can conclude that 
the photoorientation is a non-equilibrium process
in a rather complex polymer system and 
it still remains a challenge
to develop a tractable theory  treating all the above points
adequately.

As far as the long term stability
of POA is concerned, 
the reorientation of the azobenzene groups can be assumed to
result in the appearance of a self-consistent anisotropic field that
support induced anisotropy.  This field comes from anisotropic
interactions between the azobenzene fragments and rearrangement of the
main chains and other non-absorbing fragments. 
In other words, the photoinduced orientational structures can
be regarded as a result of the photo-reorientation and 
thermotropic self-organi\-za\-tion
processes~\cite{Stu:1997a}.

There are two phenomenological models based on similar assumptions:
the multidomain model proposed in~\cite{Ped:1997} and the model with
additional order parameter, attributed to the polymer backbone and
introduced to make the steady state degenerate~\cite{Puch:1998}. 

Despite these models look different it is clear that they
incorporate the long term stability by introducing additional
degree of freedom (subsystem) which kinetics would reflect cooperative
motion and account for non-equilibrium behavior. 

In this section we describe our theoretical approach to
the kinetics of the photoinduced reorientation.
We begin with the analysis of general master equations and 
specialize then the rates of the involved transitions.
The resulting kinetic equations for order parameters 
are derived after making assumptions
on the form of  angular redistribution probabilities
and the order parameter correlation functions. 
In addition, we show that this approach can be employed to 
derive the known phenomenological
models~\cite{Ped:1997,Puch:1998,Puch:1999}.
Finally at the end of the section we present
the numerical results obtained by solving the kinetic equations
for the order parameters and concentrations.

\subsection{Master equations}
\label{subsec:master}

We shall assume that the dye molecules in 
the ground state  are of \trans form 
with the orientation of the molecular axis defined by
the unit vector $\uvc{n}$. The latter is specified by
the polar, $\theta$, and azimuthal, $\phi$, angles:
$\uvc{n}=$($\sin\theta\cos\phi$,$\sin\theta\sin\phi$,
$\cos\theta$). 

Angular distribution of the \trans molecules at time $t$
is characterized by the number distribution function $N_{tr}(\uvc{n},t)$.
Molecules in the excited state have the \cis conformation and
the corresponding function is $N_{cis}(\uvc{n},t)$.
Then for the number of \trans and \cis molecules we have
\begin{gather}
  N_{tr}(t)\equiv N n_{tr}(t)=\int N_{tr}(\uvc{n},t)\,\dd\uvc{n},
  \label{eq:1t}\\
  N_{cis}(t)\equiv N n_{cis}(t)=\int N_{cis}(\uvc{n},t)\,\dd\uvc{n},
  \label{eq:1c}\\
N=N_{tr}(t)+N_{cis}(t)\, ,\qquad n_{tr}(t)+n_{cis}(t)=1\, ,
  \label{eq:consv}
\end{gather}
where 
$\displaystyle
\int\,\dd\uvc{n}\equiv 
\int_{0}^{2\pi}\dd\phi\int_{0}^{\pi}\,\sin\theta\,\dd\theta\,$ 
and 
$N$ is the total number of molecules.

We shall refer the additional subsystem 
that is able to accumulate induced ordering of the side chain molecules 
as a polymer system (matrix).
From the phenomenological point of view,
this system can be thought to represent some collective degrees of freedom
of non-absorbing units such as main chains. We shall suppose that
it is characterized  by the angular distribution function
$f_p(\uvc{n},t)$, so that $N_p(\uvc{n},t)=N_p f_p(\uvc{n},t)$.
Note that the coefficient $N_p$ can be considered as an effective
number of the units related to the polymer system.
But, more precisely, this factor determine the relations between
different thermal relaxation constants (see Eq.~\eqref{eq:cond-eq}). 

The starting point of our approach is
the kinetic rate equations taken in the following
general form~\cite{Kamp,Gard}:

\begin{align}
    \pdr{N_\alpha}&=\left[\drf{N_\alpha}\right]_{\mathrm{Diff}}+
\sum_{\beta\ne\alpha}\int
\Bigl[\,W_{\alpha\leftarrow\beta}(\uvc{n},\uvc{n}')\,
N_\beta(\uvc{n}',t)-\notag\\
&-W_{\beta\leftarrow\alpha}(\uvc{n}',\uvc{n})\,N_\alpha(\uvc{n},t)\,
\Bigr]\,\dd\uvc{n}'\, ,\quad
\label{eq:master}
\end{align}
where $\alpha\,, \beta\in\{tr,\,cis,\,p\}$.

The first term on the right hand side of Eq.~(\ref{eq:master}) is due to
rotational diffusion of molecules in \trans ($\alpha=tr$) and \cis 
($\alpha=cis$)
conformations. Note that the terms proportional to 
$W_{\alpha\leftarrow\alpha}$ can be incorporated into this diffusion
term. In what follows it is supposed that
\begin{equation}
  \label{eq:diff}
  \left[\drf{N_{tr}}\right]_{\mathrm{Diff}}=
\left[\drf{f_p}\right]_{\mathrm{Diff}}=0\,.
\end{equation}

Now
in order to proceed we need to specify the rates of the transitions.

\subsection{Transition rates}
\label{subsec:rates}

The \emph{trans--cis} transition is
stimulated by the incident UV--light quasiresonant to the
transition. Assuming that the electromagnetic 
wave is linearly
polarized along the $x$--axis, the transition rate
can be written as follows~\cite{Dum:1992,Dum:1996}:
\begin{gather}
W_{cis\leftarrow tr}(\uvc{n},\uvc{n}')=
\Gamma_{c-t}(\uvc{n},\uvc{n}')\,P_{tr}(\uvc{n}'),\notag\\
\int\Gamma_{c-t}(\uvc{n},\uvc{n}')\,\dd\uvc{n}=1\, ,
  \label{eq:wc-t}\\
\begin{align}
P_{tr}(\uvc{n}')&=(\hbar\omega_t)^{-1}\Phi_{tr\to cis}\sum_{i,j}
\sigma_{ij}^{(tr)}(\uvc{n}')E_i E_j^{*}=\notag\\
&=q_t I (1+u\, n_x^2)\equiv
q_t I (1+u\,(2S_x+1)/3)\, ,
\label{eq:ptr}
\end{align}
\end{gather}
where 
$\bs{\sigma}^{(tr)}(\uvc{n})$ is the tensor of 
absorption cross section
for the \trans molecule oriented along $\uvc{n}$:
${\sigma}^{(tr)}_{ij}=\sigma_{\perp}^{(tr)}\delta_{ij}+
(\sigma_{||}^{(tr)}-\sigma_{\perp}^{(tr)})\,
n_i\,n_j$;
$
u\equiv
(\sigma_{||}^{(tr)}-\sigma_{\perp}^{(tr)})/\sigma_{\perp}^{(tr)}
$
is the absorption anisotropy parameter;
$\hbar\omega_t$ is the photon energy;
$\Phi_{tr\to cis}$ is the quantum yield of the process and 
$\Gamma_{tr}(\uvc{n},\uvc{n}')$ describes the angular redistribution of
the molecules excited in the \cis state;
$I$ is the pumping intensity.
 
Similar line of reasoning applies
to the \emph{cis--trans} transition, so we have
\begin{align}
  \label{eq:wt-c}
& W_{tr\leftarrow cis}(\uvc{n},\uvc{n}')=
\gamma_c\Gamma_{t-c}^{(sp)}(\uvc{n},\uvc{n}')+
q_c I\,\Gamma_{t-c}^{(ind)\,}(\uvc{n},\uvc{n}'),\notag\\
& q_c\equiv \Phi_{cis\to trans}\sigma^{(cis)}\, ,
\end{align}
where $\gamma_c\equiv 1/\tau_c$, $\tau_c$ is the lifetime of \cis
molecule and the anisotropic part of the absorption cross section is
disregarded,
$\sigma_{||}^{(cis)}=\sigma_{\perp}^{(cis)}\equiv\sigma^{(cis)}$.
Eq.~\eqref{eq:wt-c} implies that the process of angular redistribution
for induced and spontaneous transitions can differ. 
Note that the normalization condition for all the angular redistribution
probability intensities is
\begin{equation}
  \label{eq:norm}
  \int\Gamma_{\alpha-\beta}(\uvc{n},\uvc{n}')\,\dd\uvc{n}=1\, .
\end{equation}

The remaining part of transitions describes equilibrating between the side
chain absorbing molecules and the polymer system. 
The corresponding rates can be taken in
the form:
\begin{align}
  \label{eq:eql}
& W_{\alpha\leftarrow p}(\uvc{n},\uvc{n}')=\gamma_{\alpha-p}
  \Gamma_{\alpha-p}(\uvc{n},\uvc{n}'),\notag\\
& W_{p\leftarrow\alpha}(\uvc{n},\uvc{n}')=\gamma_{p-\alpha}
\Gamma_{p-\alpha}(\uvc{n},\uvc{n}'),\quad \alpha\in\{tr,\,cis\}\, ,
\end{align}
where $\gamma_{\alpha-p}$ and $\gamma_{\alpha-p}$ are 
angular independent. In addition, since thermal relaxation does not
change the number of molecules in a particular state we have
the relation for $\gamma_{\alpha-p}$ and $\gamma_{p-\alpha}$:
\begin{equation}
  \label{eq:cond-eq}
  N_p \gamma_{\alpha-p}=N n_\alpha \gamma_{p-\alpha}\, .
\end{equation}
As it was mentioned above, this equation relates the thermal relaxation
constants of the polymer and the fragments through
the coefficient $N_p$, introduced in Sect.~\ref{subsec:master}.

\subsection{Model}
\label{subsec:model}

At this stage it is convenient to introduce normalized angular
distribution functions, $f_\alpha(\uvc{n},t)$:
\begin{equation}
  \label{eq:dis}
  N_\alpha(\uvc{n},t)=N n_\alpha(t) f_\alpha(\uvc{n},t)\, . 
\end{equation}

From Eqs.~\eqref{eq:master},~\eqref{eq:wc-t} and~\eqref{eq:wt-c} it is
not difficult to obtain equation for $n_{tr}(t)$:
\begin{equation}
  \label{eq:n-tr}
  \pdr{n_{tr}(t)}=
(\gamma_c+q_c I)\,n_{cis}(t)-\langle P_{tr}\rangle_{tr} n_{tr}(t)\, ,
\end{equation}
where the angular brackets $\langle\ldots\rangle_\alpha$ 
stand for averaging over angles with the distribution function
$f_{\alpha}$ . Owing to the condition~\eqref{eq:norm}, this
equation does not depend on the form of the angular redistribution
probabilities.

From the results of the previous section 
and Eq.~\eqref{eq:n-tr} we derive the equations for the
distribution functions
\begin{align}
& n_{cis}(t)\,\pdr{f_{cis}(\uvc{n},t)}= n_{cis}(t)\Biggl\{
\left[\drf{f_{cis}}\right]_{\mathrm{Diff}}-\notag\\
&-\gamma_{cis}\,f_{cis}(\uvc{n},t)
\Biggr\}-\langle P_{tr}\rangle_{tr} n_{tr}(t)+\notag\\
&+n_{tr}(t)
\int\Gamma_{c-t}(\uvc{n},\uvc{n}')P_{tr}(\uvc{n}')
f_{tr}(\uvc{n}',t)\,\dd\uvc{n}'+\notag\\
&+\gamma_{cis}\,n_{cis}(t)
\int\Gamma_{c-p}(\uvc{n},\uvc{n}')
f_{p}(\uvc{n}',t)\,\dd\uvc{n}'\, ,
\label{eq:gen-cis}
\end{align}
\begin{align}
& n_{tr}(t)\,\pdr{f_{tr}(\uvc{n},t)}= -n_{tr}(t)
\left[ P_{tr}(\uvc{n})-\langle P_{tr}\rangle_{tr}
+\gamma_{tr}\right] f_{tr}(\uvc{n},t)+
\notag\\
&+\gamma_c\,n_{cis}(t)
\int\Gamma_{t-c}^{(sp)}(\uvc{n},\uvc{n}')
f_{cis}(\uvc{n}',t)\,\dd\uvc{n}'+\notag\\
&+q_c I n_{cis}(t)
\int\Gamma_{t-c}^{(ind)}(\uvc{n},\uvc{n}')
f_{cis}(\uvc{n}',t)\,\dd\uvc{n}'-
\notag\\
&-(\gamma_c+q_c I)\,n_{cis}(t)\,f_{tr}(\uvc{n},t)+\notag\\
&+\gamma_{tr}\,n_{tr}(t)
\int\Gamma_{t-p}(\uvc{n},\uvc{n}')
f_{p}(\uvc{n}',t)\,\dd\uvc{n}'\, ,
\label{eq:gen-tr}
\end{align}
\begin{align}
&\pdr{f_{p}(\uvc{n},t)}= -\gamma_p^{(tr)}\,n_{tr}(t)
\Biggl(f_p(\uvc{n},t)-\notag\\
&-\int\Gamma_{p-t}(\uvc{n},\uvc{n}')
f_{tr}(\uvc{n}',t)\,\dd\uvc{n}'\Biggr)-
\gamma_p^{(cis)}\,n_{cis}(t)
\Biggl(f_p(\uvc{n},t)-\notag\\
&-\int\Gamma_{p-c}(\uvc{n},\uvc{n}')
f_{cis}(\uvc{n}',t)\,\dd\uvc{n}'\Biggr)\, ,
\label{eq:gen-p}
\end{align} 
where $\gamma_{tr}\equiv\gamma_{p-trans}$,
$\gamma_{cis}\equiv\gamma_{p-cis}$ and
$\gamma_{p}^{(\alpha)}\equiv N \gamma_{\alpha}/N_p$.
These equations supplemented with Eq.~\eqref{eq:n-tr} are derived
on the basis of quite general considerations. 
They can be regarded as a starting point for the formulation
of a number of phenomenological models.
We can now describe our model.

Our basic assumptions of the angular redistribution 
operators $\Gamma_{\alpha-\beta}$ are as follows
\begin{align}
 &\gamma_{cis}=0,\quad
 \Gamma_{t-c}^{(sp)}(\uvc{n},\uvc{n}')=
\Gamma_{t-p}(\uvc{n},\uvc{n}')=\notag\\
&\phantom{\gamma_{cis}=0,\quad
 \Gamma_{t-c}^{(sp)}(\uvc{n},\uvc{n}')} 
=\Gamma_{p-t}(\uvc{n},\uvc{n}')=\delta(\uvc{n}-\uvc{n}'),
 \label{eq:as-sp}\\
&\Gamma_{t-c}^{(ind)}(\uvc{n},\uvc{n}')=f_{tr}(\uvc{n},t),\quad
\Gamma_{c-t}(\uvc{n},\uvc{n}')=f_{cis}(\uvc{n},t)\, .
\label{eq:as-ind}
\end{align}

It gives the resulting system of kinetic equations:
\begin{subequations}
\label{eq:mod}
\begin{align}
& \pdr{f_{cis}}= 
\left[\drf{f_{cis}}\right]_{\mathrm{Diff}}
\label{eq:mod-cis}\\
& n_{tr}\pdr{f_{tr}}=\left(
\langle P_{tr}\rangle_{tr}-P_{tr}-\gamma_{tr}
\right)n_{tr} f_{tr}+\notag\\ 
&+\gamma_c\,n_{cis}\,(\,f_{cis}-f_{tr}\,)+\gamma_{tr}\,n_{tr}\,f_p\, ,
\label{eq:mod-tr}\\
&\pdr{f_{p}}=\gamma_p\, n_{tr}(f_{tr}-f_p)\, , 
\label{eq:mod-p}
\end{align}
\end{subequations}
where $\gamma_p\equiv\gamma_p^{(tr)}$.

\begin{figure*}[!tbh]
\centering
\subfigure[$t=0.0$~min]{%
\resizebox{85mm}{!}{\includegraphics{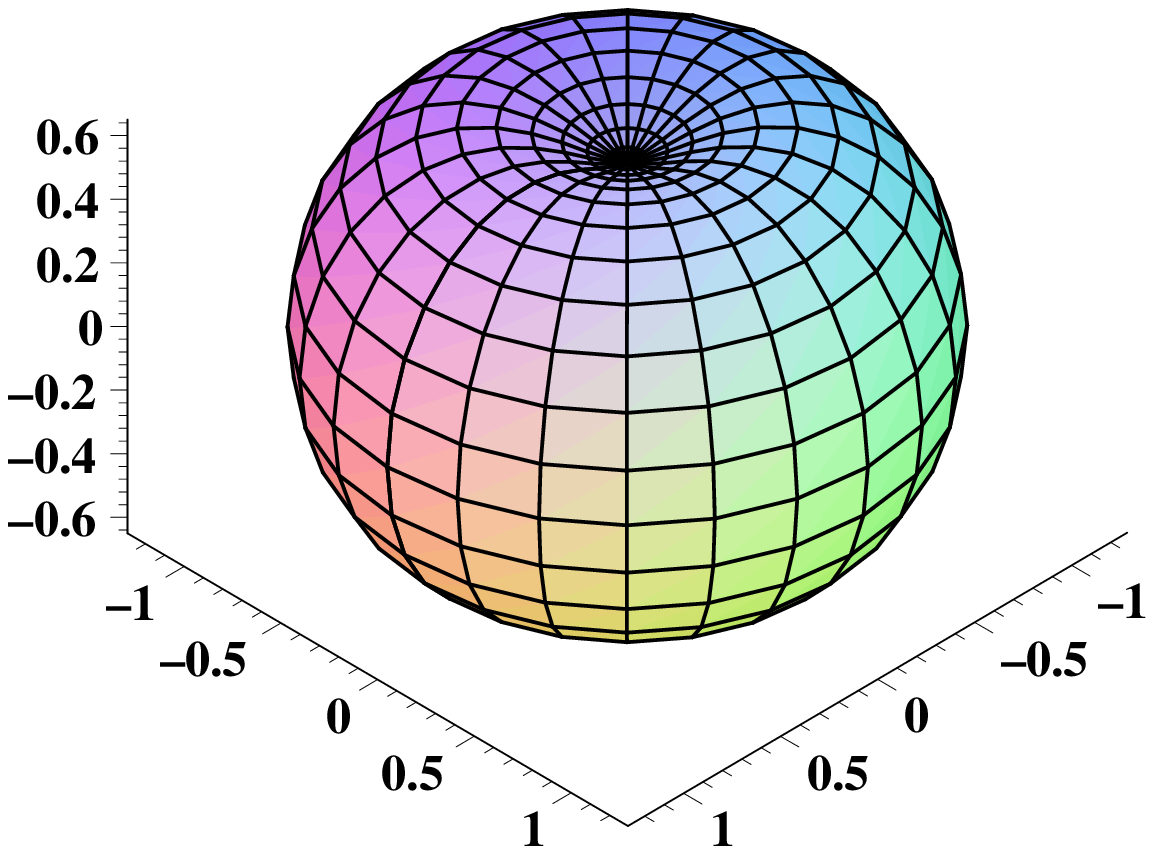}}}
\subfigure[$t=10.0$~min]{%
\resizebox{85mm}{!}{\includegraphics{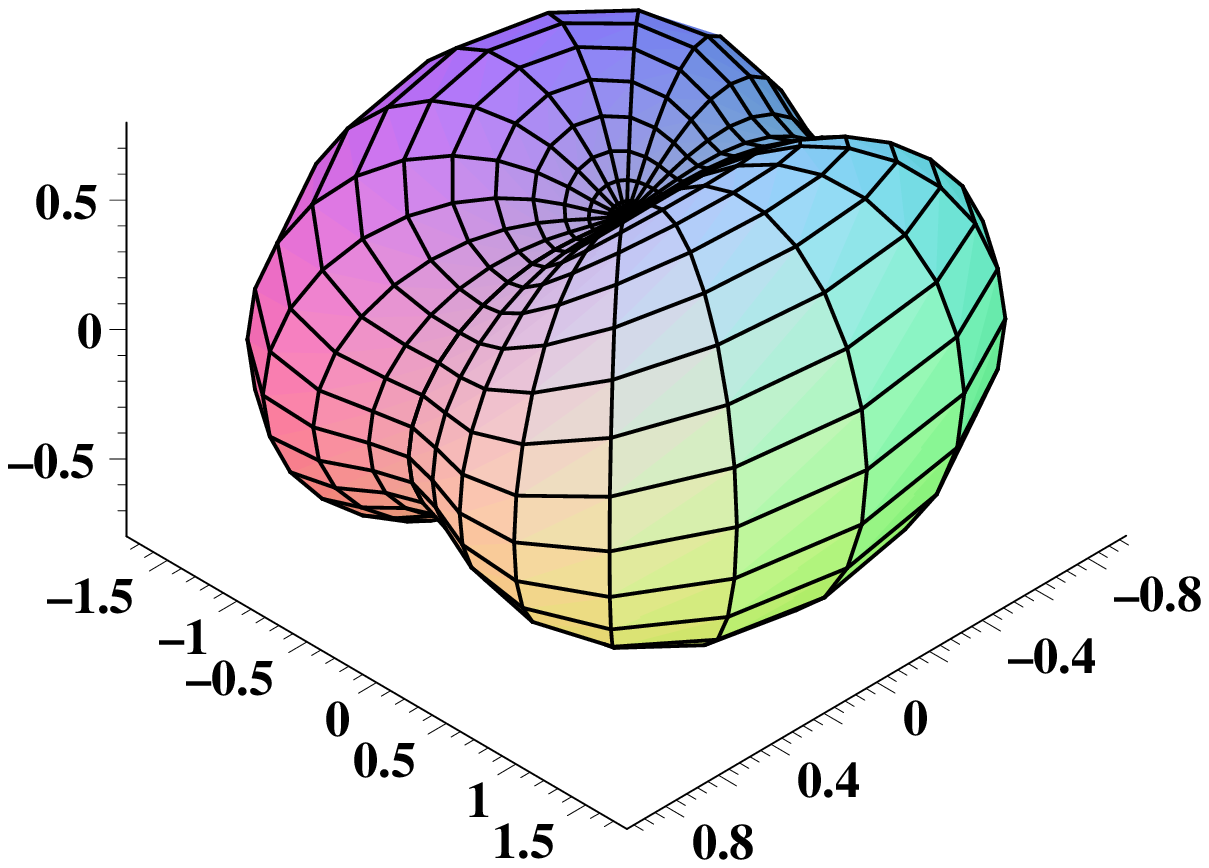}}}
\subfigure[$t=120.0$~min]{%
\resizebox{85mm}{!}{\includegraphics{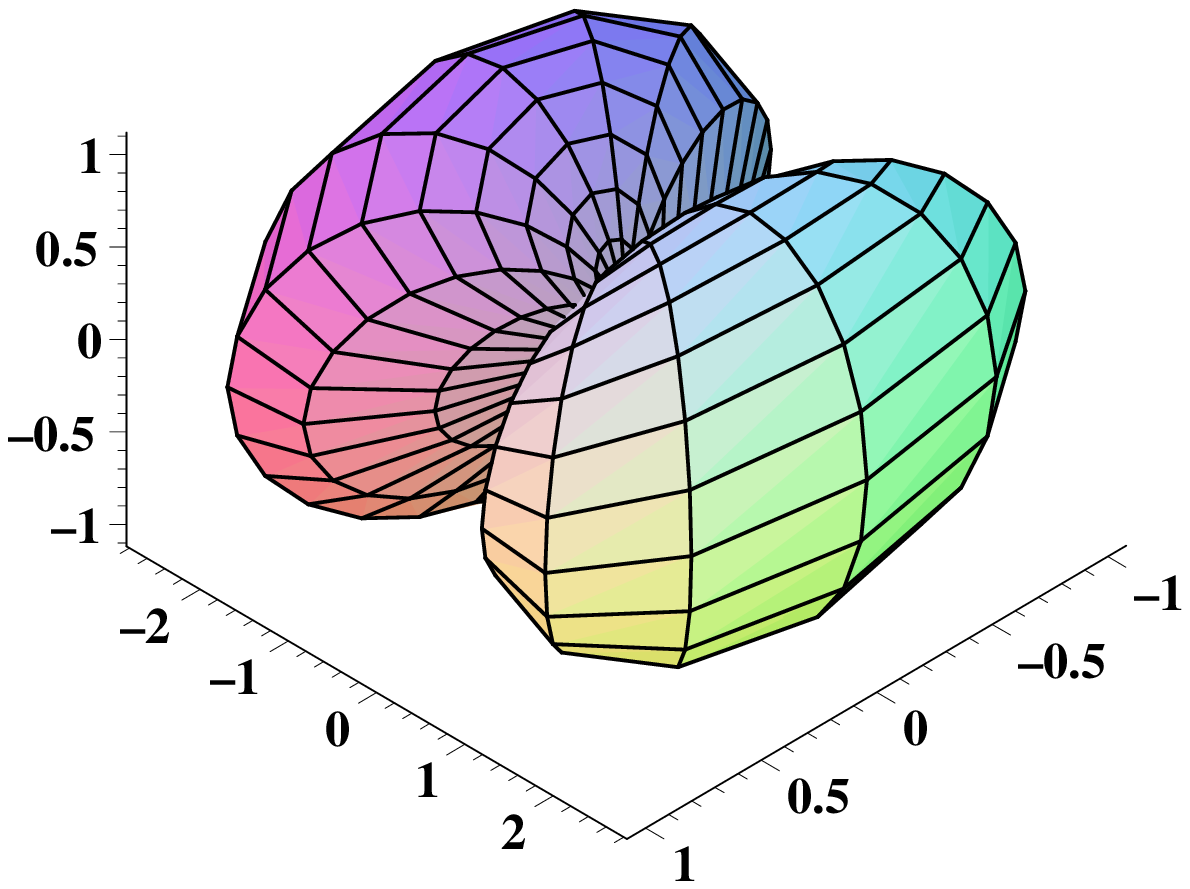}}}
\caption{%
The orientational distribution functions of \trans
molecules at different illumination doses.
These surfaces are defined
by the equation:
$r(\theta.\phi)=4\pi\,n_{tr}(t)\,f_{tr}(\theta,\phi,t)$
in the spherical coordinate system at irradiation
time $t$.
}
\label{fig:2}
\end{figure*}

Clearly, the meaning of Eq.~\eqref{eq:as-sp} is that
the molecules do not change orientation under spontaneous
transitions. On the other hand, from Eq.~\eqref{eq:as-ind}
projecting onto the angular distribution function of the corresponding
state describes angular redistribution for the stimulated transitions.

In order to explain the meaning of the projectors, note that the
multidomain model considered in~\cite{Ped:1997} can be derived from
Eqs.~\eqref{eq:gen-cis}~--~\eqref{eq:gen-p} 
by putting $\gamma_c=0$, $\gamma_{cis}=\gamma_{tr}$
and assuming that all angular redistribution probabilities are equal
to the equilibrium distribution,
$\Gamma_{\alpha-\beta}(\uvc{n},\uvc{n}')=p(\uvc{n})$, determined by
the mean field potential $W(\uvc{n})$: $p(\uvc{n})\propto \exp(-W)$.
In other words, this procedure introduces the mean field potential by
assuming that the angular redistribution operators
$\Gamma_{\alpha-\beta}$ act as projectors onto the equilibrium
distribution.
Note that the results of~\cite{Puch:1998,Puch:1999}  
correspond to the case where
$\Gamma_{t-c}^{(sp)}(\uvc{n},\uvc{n}')=f_{tr}(\uvc{n},t)$
and $\Gamma_{t-c}^{(ind)}(\uvc{n},\uvc{n}')=(4\pi)^{-1}$.

\subsection{Order parameters}
\label{subsec:ord-par}

We can now deduce the equations that describe the temporal evolution of
the diagonal components of the order parameter tensor
\begin{equation}
  \label{eq:tens}
  \vc{S}(\uvc{n})=2^{-1}\,(3 n_i n_j -\delta_{ij})\,
\uvc{e}_i\otimes\uvc{e}_j\, .
\end{equation}

The components of interest can be expressed in terms of
Wigner $D$--functions~\cite{Bie} as follows
\begin{align}
& S_x=(3 n_x^2-1)/2=\notag\\
&\phantom{S_x}
=-(D_{00}^2(\uvc{n})-\sqrt{6}\,\Re D_{20}^2(\uvc{n}))/2\,,
\label{eq:sx}
\end{align}
\begin{align}
& S_y=(3 n_y^2-1)/2=\notag\\
&\phantom{S_y}
-(D_{00}^2(\uvc{n})+\sqrt{6}\,\Re D_{20}^2(\uvc{n}))/2\,,
\label{eq:sy}
\end{align}
\begin{equation}
S_z=(3 n_z^2-1)/2=
D_{00}^2(\uvc{n})\, .
\label{eq:sz}
\end{equation}

The simplest case occurs for the order parameters of \cis molecules.
Eq.~\eqref{eq:mod-cis} yields the following result:
\begin{equation}
  \label{eq:ord-cis}
  \pdr{S_{ij}^{(cis)}}=-6 D_r S_{ij}^{(cis)}\, ,
\end{equation}
where $ S_{ij}^{(\alpha)}\equiv\langle S_{ij}(\uvc{n})\rangle_\alpha$ and
$D_r$ is the rotational diffusion constant. Clearly, our assumptions
correspond to the case where the presence of \cis molecules is of
minor importance for ordering kinetics.

Eqs.~\eqref{eq:ptr},~\eqref{eq:mod-tr} and~\eqref{eq:mod-p}
give the following system for the components of the order parameter
tensor: 
\begin{subequations}
\label{eq:gen-ord}
\begin{align}
&n_{tr}\pdr{S_{ij}^{(tr)}}= -2/3\, q_t I u\,
n_{tr}\,G_{ij;\,xx}^{(tr)}+\notag\\
&+\gamma_c n_{cis} (S_{ij}^{(cis)}-S_{ij}^{(tr)}) +
\gamma_{tr} n_{tr} (S_{ij}^{(p)}-S_{ij}^{(tr)})\, ,
\label{eq:gen-ord-tr}\\
&\pdr{S_{ij}^{(p)}}=-\gamma_{p} n_{tr}(S_{ij}^{(p)}-S_{ij}^{(tr)})\, ,
\label{eq:gen-ord-p}
\end{align}
\end{subequations}
where $G_{ij;\,mn}^{(\alpha)}$ is the correlation function of the
order parameter components $S_{ij}(\uvc{n})$ and $S_{mn}(\uvc{n})$
that is defined by the following relation
\begin{equation}
  \label{eq:cor-fun}
  G_{ij;\,mn}^{(\alpha)}=
\langle S_{ij}(\uvc{n})S_{mn}(\uvc{n}) \rangle_\alpha-
S_{ij}^{(\alpha)}\,S_{mn}^{(\alpha)}\, .
\end{equation}

Computing the order parameter correlation functions that enter
Eqs.~\eqref{eq:gen-ord} requires the knowledge of details on
microscopic interactions and, in general, for nonequilibrium system it
can be rather involved and sophisticated.  In this paper, 
we shall adopt the simplest ``kinematic'' procedure to
approximate the correlators. 
It implies that after writing the products of $D$--functions as a
sum of spherical harmonics we neglect the high order harmonics with
angular momentum $j>2$. In particular, we have
\begin{equation}
  \label{eq:decoup}
  \langle S_\alpha^2\rangle_{tr}\approx
1/5+2/7\,\langle S_\alpha\rangle_{tr}\, ,\quad
\alpha\in\{x,\,y,\,z\}\, .
\end{equation}

Applying this procedure to Eqs.~\eqref{eq:gen-ord} leads to the result
given by
\begin{align}
&n_{tr}\pdr{S}= 2u/3\, q_t I (1/5+2/7\,S-S^2) n_{tr} -\notag\\
&-\gamma_c n_{cis} S +
+\gamma_{tr} n_{tr} (S_p-S)\, ,
\label{eq:ord-tr1}
\end{align}
\begin{align}
&n_{tr}\pdr{\Delta S}= -2u/3\, q_t I (2/7+S) n_{tr}\Delta S -\notag\\
&-\gamma_c n_{cis} \Delta S +
\gamma_{tr} n_{tr} (\Delta S_p-\Delta S)\, ,
\label{eq:ord-tr2}
\end{align}
\begin{align}
&\pdr{S_p}=-\gamma_{p} n_{tr} (S_p-S)\, ,
\label{eq:ord-p1}
\end{align}
\begin{align}
&\pdr{\Delta S_p}=-\gamma_{p} n_{tr} (\Delta S_p-\Delta S)\, ,
\label{eq:ord-p2}
\end{align}
where $S\equiv \langle S_y\rangle_{tr} $,
$\Delta S\equiv \langle S_x-S_z\rangle_{tr} $,
$S_p\equiv \langle S_y\rangle_{p} $ and
$\Delta S_p\equiv \langle S_x-S_z\rangle_{p} $.
Eqs.~\eqref{eq:ord-tr1} --~\eqref{eq:ord-p2} combined with
Eq.~\eqref{eq:n-tr} form the system of kinetic equations
for our phenomenological model.

\begin{figure}[!tbh]
\vskip2mm
\centering
\resizebox{70mm}{!}{\includegraphics{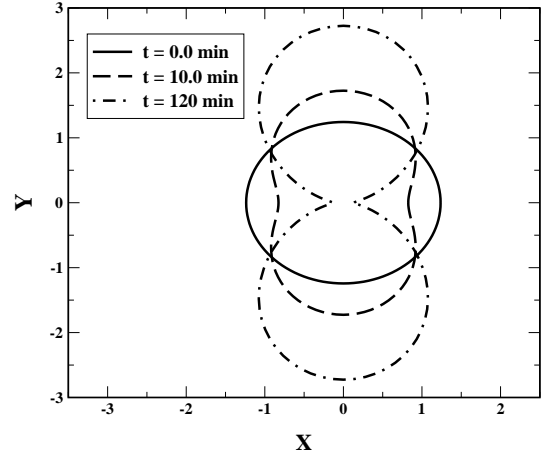}}
\caption{%
Initial orientational distribution at
section by the $x$-$y$ plane is isotropic. 
It assumes anisotropy along the $y$ axis
as illumination time increases.
}
\label{fig:3}
\end{figure}

\subsection{Numerical results}
\label{subsec:num-res}

Theoretical curves depicted in Fig.~\ref{fig:1}b are calculated
by solving the equations deduced in the previous section.
This procedure involves computing the stationary values of
$S$ and $\Delta S$ to which the order parameters decays
after switching off the irradiation at time $t$.
In addition, we need to take into consideration the difference between the
order parameters defined by Eq.~\eqref{eq:3} and
the order parameters of Sect.~\ref{subsec:ord-par}.
Since $D_i\propto (1+u(2\,S_i+1)/3)$, these order parameters differ
by the factor $u/(3+u)$.

According to the experimental data, the lifetime of \cis
molecules $\tau_c$ ($\gamma_c=1/\tau_c$) is about 1.0~sec,
whereas the relaxation time after switching off the irradiation
can be estimated at  4~min. 
Since the theoretical value of this relaxation time is
$1/(\gamma_p + \gamma_{tr})$,
the relaxation times $\tau_p$ ($\gamma_p=1/\tau_p$) and $\tau_{tr}$
($\gamma_{tr}=1/\tau_{tr}$) can be taken to be equal 8~min.

We estimated the absorption cross sections 
$\sigma^{(cis)}$ and $\sigma^{(tr)}$ from the UV spectra
of the polymer solved in toluene. These spectra were measured 
before and during irradiation. In the latter case 
the solution was in the photosaturated state. The absorption bands were
then decomposed into the bands of \trans and \cis isomers to
yield the corresponding values of the extinction coefficients
at $\lambda_t=365$~nm. In order to compute these coefficients we 
followed the procedure described in~\cite{Bern:1975}.  
The resulting estimates can be written as follows: 
$(\hbar\omega_t)^{-1}\sigma^{(cis)}\tau_c I\approx 0.2\times 10^{-2}$ 
and 
$\sigma^{(tr)}/\sigma^{(cis)}\approx 2.5$ for
$I=1$~mW/cm$^2$, where
$\sigma^{(tr)}=(\sigma_{||}^{(tr)}+2\sigma_{\perp}^{(tr)})/3$
is the average absorption cross section of the \trans fragments.

Then given the quantum yield $\Phi_{cis\to trans}$
and the experimental value of 
the order parameter in the photosteady state, $S_{st}=0.251\,(1+3/u)$,
we can compute the anisotropy parameter $u$ from the
equation for $S_{st}$. The latter
can be derived from Eqs.~\eqref{eq:n-tr}
and~\eqref{eq:ord-tr1} 
by setting the time derivatives on the left hand sides equal to zero:
\begin{align}
\gamma_c S_{st}&(1+u(1-S_{st})/3)=\notag\\
&=2 u/3\, (\gamma_c+q_c I)\,(1/5\,+2/7\, S_{st}-S_{st}^{\,2})\, .
  \label{eq:st}
\end{align}

The theoretical curves in Fig.~\ref{fig:1}b are calculated at
$\Phi_{cis\to trans}=10$\% and $\Phi_{cis\to trans}=5$\%
that is to yield the value of the ratio
$\sigma_{||}^{(tr)}/\sigma_{\perp}^{(tr)}=8.9$.
Note that the quantum efficiencies are of the same order of magnitude as
the experimental values for other azobenzene
compounds~\cite{Mita:1989}. On the other hand, we have
$q_c\approx 0.2\times 10^{-3}$~cm$^2$/mJ that is about the value
given in~\cite{Ped:1997}.

The computed order parameter components can be used to illustrate
orientational distributions of \trans fragments maintained after
different illumination doses.  
Fig.~\ref{fig:2} shows the surfaces
$r(\theta.\phi)=4\pi\,n_{tr}(t)\,f_{tr}(\theta,\phi,t)$ that indicate
the angular redistribution in the course of irradiation. 
Note that we have truncated  
the expansion for the distribution function $f_{tr}$
by neglecting the high order spherical harmonics:
\begin{align}
4\pi\,f_{tr}(\uvc{n},t)\approx &
1+5\,\Bigl[
\langle S_z\rangle_{tr} D^{\,2}_{00}(\uvc{n})+\notag\\
&+\frac{\langle S_x\rangle_{tr}-\langle S_y\rangle_{tr}}{\sqrt{6}}\,
2\,\Re D^{\,2}_{20}(\uvc{n})
\Bigr]\, .
  \label{eq:fig-f_tr}
\end{align}

\begin{figure}[!tbh]
\vskip5mm
\centering
\resizebox{70mm}{!}{\includegraphics{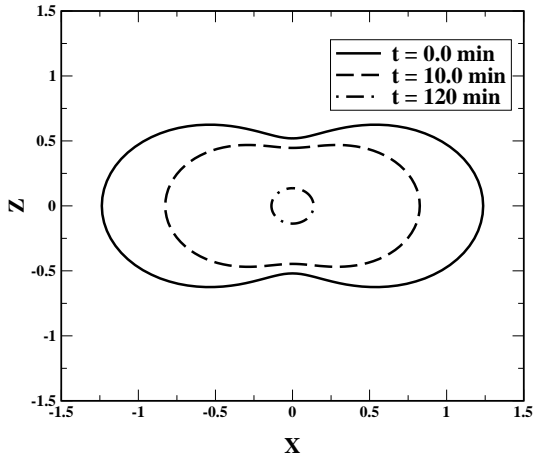}}
\caption{%
Increase in the illumination dose renders
the orientational distribution of the \trans
fragments at section by the $x$-$z$ plane
isotropic.
}
\label{fig:4}
\end{figure}

Sections of the surfaces depicted in Fig.~\ref{fig:2} by
the $x$-$y$ and $x$-$z$ coordinate planes are shown
in Figs.~\ref{fig:3}~--~\ref{fig:4}.
As is seen from Fig.~\ref{fig:3}, the angular distribution in
the $x$-$y$ plane becomes anisotropic under the action of light,
whereas Fig.~\ref{fig:4} indicates that the distribution
in the $x$-$z$ plane goes isotropic. 

\section{Conclusions}
\label{sec:concl}

In this paper we demonstrate that the combination of
the UV/Vis absorption spectroscopy and the suitably modified
null ellipsometry method is a tool appropriate
for a comprehensive study of POA in films
of liquid crystalline polymers with azobenzene side groups.

We found that initially the spincoated films under
investigation are characterized by 
a preferred in-plane orientation of the azobenzene fragments, whereas
the long axes are randomly distributed in this plane. 
So, the resulting
structure is isotropic in this plane, but in the space it is uniaxial
with the optical axis normal to the film surface resulting in a negative
birefringence (see Sect.~\ref{subsec:exp-res}).
 
Increasing irradiation doses results in
an anisotropic order maintained in the film after switching off 
the linearly polarized UV light.
This photo-induced structure 
corresponds to a biaxial orientational order of 
the azobenzene moieties.
But it becomes uniaxial at sufficiently large doses
reaching the photosaturated state of the photoorientation process.
The process results usually in an oblate order in the case of
amorphous polymers.
The uniaxiality in the studied case 
could be related to liquid crystalline character of the polymer or
the initial order of the film which act in combination with the
linearly polarized light as aligning force resulting in an uniaxial
in-plane order.
So, the action of the actinic light can be regarded as a factor
stimulating both photoorientation and 
thermotropic self-organization.

Quantitatively, experimental results on the kinetics of
the photoreorientation were described
in terms of the order parameters 
introduced in Sect.~\ref{subsec:exp-res}.
It makes the comparison between
experimental and theoretical results relatively direct.
On the other hand, it raises the question as to the validity of the
procedure used to determine the out-of-plane absorption
$D_z$. In our case this can be justified for
the number of \trans fragments is shown to remain unchanged
in all the relaxed states caused by
the lifetime of \cis isomer.
So, the conclusion is that
the photoreorientation in the film  occurs through the mechanism of 
angular redistribution.
Note that,
as a matter of fact, the latter was implicitly assumed in our
theoretical model formulated in Sect.~\ref{subsec:model}.
 
Despite the theoretical considerations of Sect.~\ref{sec:theory} are
rather phenomenological they emphasize the key points that should be
addressed by such kind of theories.  These are the angular
redistribution probabilities (Sect.~\ref{subsec:model}) and the order
parameter correlation functions (Sect.~\ref{subsec:ord-par}).  The
redistribution operators, in particular, define how the system relaxes
after switching off the irradiation and can serve to introduce
self-consistent fields.  The correlators, roughly speaking, mainly
determine the character of photoreorientation and the properties of
the photosaturated state.

The assumptions taken in this paper give the model with the kinetics
governed by the mechanism of angular redistribution. The form of
the approximate correlators influences it in such a way that
the out-of-plane reorientation is appeared to be suppressed.

Our simple model depends on a few 
parameters that enter the equations and that can be estimated from
the experimental data. 
Only the anisotropy
parameter and the quantum yields are derived by making comparison
between the experimental data and the theoretical dependencies. 

This theory predicts that the fraction of \cis isomers 
is negligible and the orientational kinetics of the \cis fragments
governed by Eq.~\eqref{eq:ord-cis} is irrelevant.
The latter is why the liquid crystalline ordering effects~\cite{Doi:1986}
that would complicate the kinetics of the \cis molecules can be safely
omitted in Eq.~\eqref{eq:ord-cis}.
Note, however, that it is rather straightforward to modify the model
for the case where the \cis fragments would 
affect the kinetics considerably. We shall extend on the subject
elsewhere.

More detailed microscopic theoretical approach is apparently beyond
the scope of the paper.  
Theoretically, it would be interesting to put this
theory into the context of ergodicity breaking
transitions~\cite{Kis:susy}. This work is under progress.

\begin{acknowledgement}
We acknowledge financial support from 
CRDF under the grant UP1--2121B.
We also thank Dr.~T.~Sergan and Prof.~J.~Kelly 
from Kent State University
for assistance with processing the data of the null ellipsometry
measurements and helpful discussions.

\end{acknowledgement}

\end{document}